\documentclass[twocolumn]{revtex4}
\usepackage[dvipdfm]{graphicx}
\usepackage{amsmath,amssymb,bm,amsthm}
\usepackage{color}
\usepackage{ulem}
\usepackage{here}

\newcommand{\nc}{\newcommand}
\nc{\rnc}{\renewcommand}
\nc{\nn}{\nonumber}

\begin{document}
\title{Entanglement prethermalization: Locally thermal but non-locally non-thermal states in a one-dimensional Bose gas}

\author{Eriko Kaminishi, Takashi Mori, Tatsuhiko N. Ikeda and Masahito Ueda}
\affiliation{Department of Physics, University of Tokyo, \\ 7-3-1 Hongo, Bunkyo-ku, Tokyo 113-0033, Japan} 
\date{\today}

\begin{abstract}
{\bf{A well-isolated system often shows relaxation to a quasi-stationary state before reaching thermal equilibrium.
Such a prethermalization \cite{Berges2004} has attracted considerable interest recently in association with closely related fundamental problems of relaxation and thermalization of isolated quantum systems \cite{ Gring2012, Langen2013, Kollar2011, Van2013}. 
Motivated by the recent experiment in ultracold atoms \cite{Gring2012}, we study the dynamics of a one-dimensional Bose gas which is split into two subsystems, and find that individual subsystems relax to Gibbs states, yet the entire system does not due to quantum entanglement. In view of recent experimental realization on a small well-defined number of ultracold atoms \cite{Serwane2011}, our prediction based on exact few-body calculations is amenable to experimental test. 
}}
\end{abstract}
\maketitle
Relaxation and thermalization have been studied over a wide range of fields \cite{Neumann1929, Srednicki1994, Tasaki1998, Popescu2006, Rigol2008, Linden2009, Goldstein2010, Sato2012}.
Especially, isolated quantum systems have recently been realized in cold atom experiments \cite{Kinoshita2006, Hofferberth2007}, giving an enormous impetus to researchers to explore such fundamental issues.
In the case of an integrable 1D Bose gas, thermalization does not take place due to the existence of as many conserved quantities as subsystem's degrees of freedom and the system stays at a quasi-stationary state \cite{,Kinoshita2006, Gring2012}.
It is conjectured that the prethermalized state can be described by a generalized Gibbs ensemble \cite{Rigol2007, Rigol2008} (see, however, Ref \cite{Kormos2013, De2014}).
In an actual system, there are nonnegligible integrability-breaking perturbations which cause the eventual thermalization of the system.
Thus, non-thermal steady states found in an idealized situation should be regarded as having a finite lifetime in a more realistic setup, so we call them ``prethermalized states''. 
Prethermalization has been experimentally observed in a 1D Bose gas which is coherently split into two parts, left and right \cite{Gring2012}, with the same phase.
Because the subsystems are spatially separated, such an initial correlation can persist non-locally.
Can such a nonlocal correlation or quantum entanglement affect prethermalization of the entire system?
We will answer this question in the affirmative.
The memory of those initial nonlocal correlations persists throughout time evolution, which is expected to significantly influence prethermalization.
We emphasize that the mechanism of entanglement prethermalization is general as we will see later, although we focus on the integrable 1D Bose gas in this paper. 
We will also propose a different experimental system which is expected to exhibit entanglement prethermalization.

\begin{figure}[t]
\begin{center}
\includegraphics[width=8cm]{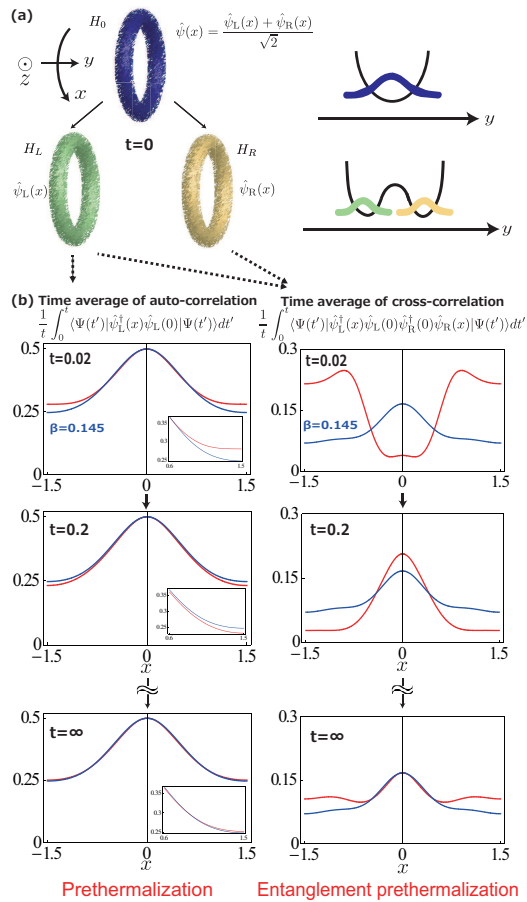}
\caption{{\bf{Schematic illustration of our study (a) and calculated correlation functions in the course of prethermalization (b). }}
{\bf{(a)}} Field operators in the Lieb-Liniger Hamiltonian before the split consist of the left(L) component and the right(R) component.
Initially, the system is tightly confined by a harmonic trap in the radial direction, so that the system is one-dimensional. 
Then, the harmonic trap is modulated into a double-well potential along the $y$-direction whose two minima give two 1D systems which we call the left and the right.
After this process, each particle at position $x$ in the original 1D system becomes a superposition of the particle on the left and right at $x$.
{\bf{(b)}} The evolution of time averages of two-point correlation functions for the number of particles $N=3$.
The cross-correlation (left-right correlation) shows entanglement prethermalization, whereas the auto-correlation (left correlation) shows thermalization at an effective temperature lower than the ``equilibrium" temperature which is determined from the energy of the system based on the canonical distribution.}
\label{setupLL}
\end{center}
\end{figure}

%
%
\begin{figure*}[t]
\begin{center}
\includegraphics[clip,width=15.0cm]{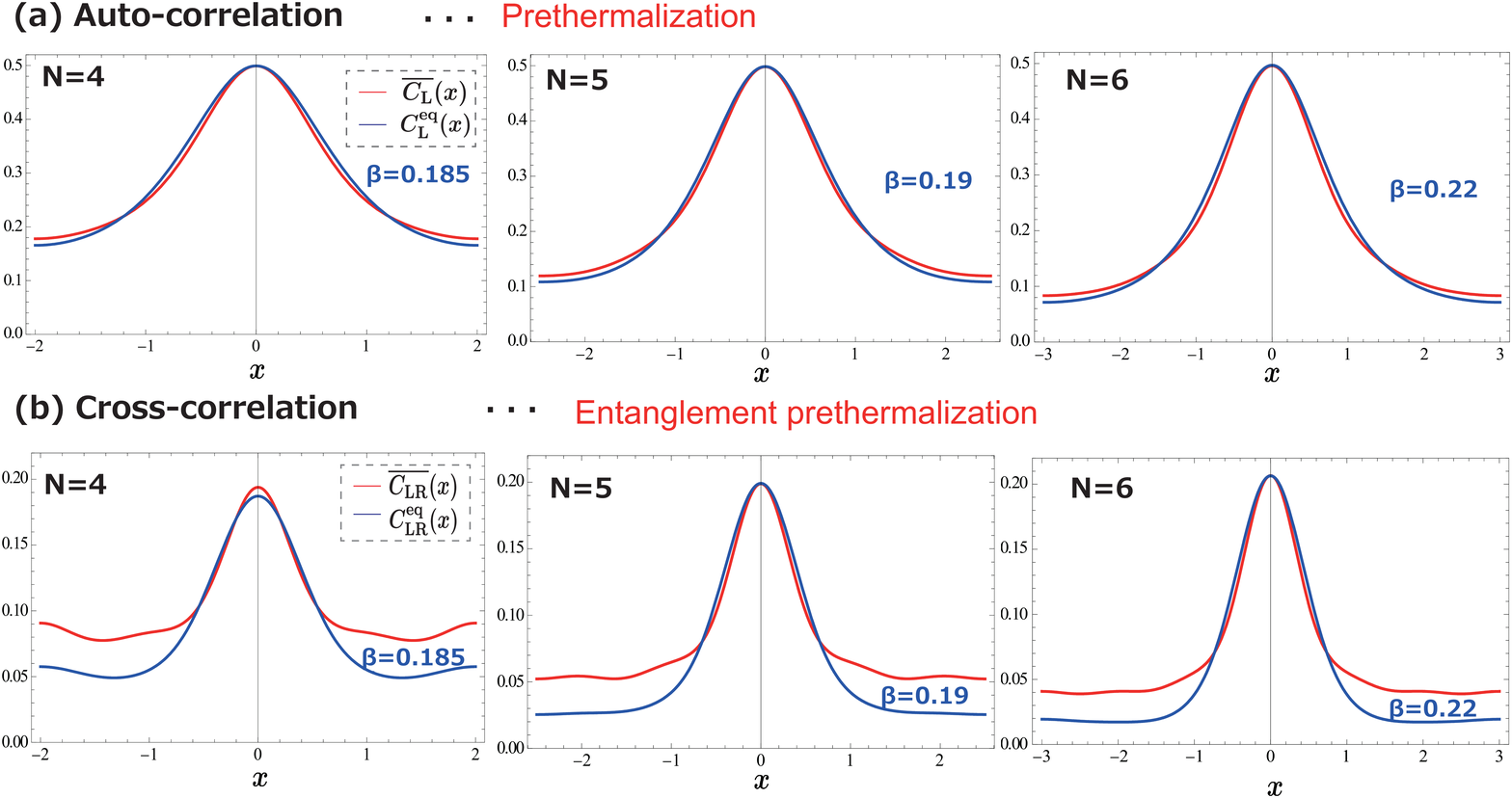}
\caption{{\bf{Prethermalization and entanglement prethermalization in two-point correlation functions.}}
{\bf{(a)}} Each auto-correlation function shows thermalization at an effective temperature $\beta^{-1}$ in the entire region.
The value of $\beta$ is determined by fitting $\overline{C_{\rm L}}(x)$ to $C_{\rm L}^{\rm eq}(x)$.
{\bf{(b)}} Each cross-correlation function shows thermalization at an effective temperature locally but no thermalization non-locally.
The red curves show infinite-time averages of the cross-correlation function $\overline{C_{\rm LR}}(x)$ calculated by the Bethe ansatz method.
The blue curves show the thermal equilibrium averages of the cross-correlation function $C_{\rm LR}^{\rm eq}(x)$ at the temperature $\beta^{-1}$. 
Non-local prethermalization is characterized by the cross-correlation function.}
\label{correlation}
\end{center}
\end{figure*}

Here, we study prethermalization in a few-body system, motivated by experimental realizations of isolated quantum systems with a few but well-defined number of particles \cite{Serwane2011}.
Theoretical analysis of the experiment \cite{Gring2012} has been done by using the Tomonaga-Luttinger theory \cite{Gring2012, Kitagawa2010, Kitagawa2011}, but it cannot be applied to a few-body system.
The difficulty of theoretically studying a few-body system is that we need to exactly solve the Schr\"odinger equation for an arbitrarily long time.
To overcome this difficulty, we adopt the Lieb-Liniger model \cite{Lieb1963}.
Although the exact eigenenergies can be obtained by solving the Bethe ansatz equations, it is challenging to calculate the time evolution of physical quantities because it requires us to evaluate the matrix elements of an observable in the basis of exact eigenstates.
A recent achievement made in Ref.~\cite{Caux2007} enables us to evaluate matrix elements of field operators by using the Gaudin-Korepin~\cite{Gaudin1983, Korepin86} and Slavnov formulae~\cite{Slavnov1989, Slavnov1990}.
By adopting this technique, we calculate the time evolution of correlation functions necessary for investigating prethermalization in split 1D Bose gases.

Before going into a detailed discussion, we briefly describe the setup of our study (see Fig.~\ref{setupLL}(a)).
We consider a 1D Bose gas which is initially prepared in the ground state $|\Psi(0)\rangle$ and split into two parts $H_{\rm L}$ and $H_{\rm R}$.
After the splitting, the two gases are allowed to evolve in time independently: $|\Psi(t)\rangle=e^{-i(H_{\rm L}+H_{\rm R})t}|\Psi(0)\rangle$.
We calculate two types of correlation functions.
One is the auto-correlation (left correlation) $C_{\rm L}(x,t)=\langle\Psi(t)|\hat{\psi}^{\dagger}_{\rm L}(x)\hat{\psi}_{\rm L}(0)|\Psi(t)\rangle$ and the other one is the cross-correlation (left-right correlation) defined by 
$C_{\rm LR}(x,t)=\langle\Psi(t)|\hat{\psi}^{\dagger}_{\rm L}(x)\hat{\psi}_{\rm L}(0)\hat{\psi}^{\dagger}_{\rm R}(0)\hat{\psi}_{\rm R}(x)|\Psi(t)\rangle$.
In a few-body system, fluctuations in time are large, so we compute the time average of these quantities $\int_0^t dt' C_{\rm L}(x,t') /t$ and $\int_0^t dt' C_{\rm LR}(x,t') /t$ which are shown in Fig.~\ref{setupLL}(b) for $N=3$.
As shown in Fig.~\ref{setupLL}(b), each auto-correlation function rapidly relaxes to an equilibrium curve at an effective temperature over the entire region.
On the other hand, the cross-correlation function approaches an equilibrium curve only at short distance but exhibits markedly different behavior at long distance.
The cross-correlation reaches a stationary value by about $t=1$.

The effective temperature determined by fitting the numerical data with the canonical distribution is found to be lower than the equilibrium temperature determined by comparing the energy of the state $|\Psi(t)\rangle$ with that of the equilibrium state; for example, $\beta=0.145$ and $\beta_{\rm{eq}}=0.134$ for $N=3$, which implies that each individual Bose gas (left or right) is in a prethermalized state.
In other words, prethermalization of an individual Bose gas after the split is described by a Gibbs state with an effective temperature.
However, prethermalization observed in the left-right interference cannot be described by any Gibbs state with some effective temperature, and hence this prethermalization is not of the same types as that observed in a single Bose gas.
The former emerges due to the initial entanglement between the left and the right as shown later; hence we call it the entanglement prethermalization.

This locally thermal but non-locally non-thermal state can also be observed for larger numbers of particles.
The infinite-time averages of the auto-correlation $\overline{C_{\rm L}}(x)$ and the cross-correlation $\overline{C_{\rm LR}}(x)$ are shown in Fig.~\ref{correlation}.
The infinite-time average of the cross-correlation function $\overline{C_{\rm LR}}(x)$ deviates at long distance from the equilibrium one $C_{\rm LR}^{\rm eq}(x)$ calculated by the equilibrium ensemble at an effective temperature (see Fig.~\ref{correlation}(b)), while the left  $\overline{C_{\rm L}}(x)$ (or right $\overline{C_{\rm R}}(x)$) correlation function is indistinguishable from the left equilibrium one $C_{\rm L}^{\rm eq}(x)$ (or right $C_{\rm R}^{\rm eq}(x)$) which is fitted by choosing an appropriate effective temperature (see Fig.~\ref{correlation}(a)).
From this observation, we conclude that the initial entanglement between the left and right persists during the many-body dynamics and prevents the system from being thermalized.
Usually, an initial entanglement is washed out in the course of time evolution and does not influence the long-time average of a physical quantity.
By extracting the effect of initial entanglement from the infinite-time average of the density matrix, 
we find that the initial entanglement can actually affect the infinite-time average of the cross-correlation function $\overline{C_{\rm LR}}(x)$.
On the other hand, we also show that the initial entanglement does not affect the long-time average of a quantity concerning the two individual Bose gases.
These facts demonstrate the entanglement prethermalization in a few-body system.

A fundamental question is how small the system can exhibit prethermalization.
To address this question, we perform exact many-body calculations for $N=3,4,5$ and 6, and find that the smallest system with $N=3$ already show not only individual thermalization but also entanglement prethermalization.
In view of a recent experimental achievement to create systems with a fixed number of atoms up to $N=10$ \cite{Serwane2011}, the result of our work should be amenable to experimental test.

Now we shall explain our study in detail.
We adopt the Lieb-Liniger model \cite{Lieb1963} which describes a 1D Bose gas with a repulsive delta-function interaction.
We impose periodic boundary conditions of the system size $L$ on the wave functions. 
The second-quantized Hamiltonian of the Lieb-Liniger model is given by
\begin{equation}
\hat{H}=\int_{0}^{L}dx\left({\partial_x}\hat{\psi}^{\dagger}(x){\partial_x}\hat\psi(x)+c \hat\psi^{\dagger}(x)\hat\psi^{\dagger}(x)\hat\psi(x)\hat\psi(x)\right),
\label{eq:LL}
\end{equation}
where $\hat{\psi}(x)$ is the bosonic field operator and we employ a system of units with $2m=\hbar=1$.
Moreover, we choose the unit of length so that the particle density is unity: $n=N/L=1$, and $c=50$.
In this model, the exact energy eigenstates can be obtained by using the Bethe ansatz method.
Each $N$-particle eigenstate $|{\bm{k}^N}\rangle$ is labeled by a set of quasi-momenta ${\bm{k}^N}\equiv\{k_j\}_{i=1}^N$ that satisfy the Bethe ansatz equations (BAEs),
\begin{align}
k_j L = 2 \pi I_j - 2 \sum_{\ell \ne j}^{N} 
\arctan \left({\frac {k_j - k_{\ell}} c } \right) ,
\label{BAE} 
\end{align}
where $j=1,2,\cdots,N$ and $I_j$'s, which are called the Bethe quantum numbers, are integers for odd $N$ and half integers for even $N$. 
The total momentum $P$ and energy eigenvalue $E$ are given in terms of quasi-momenta as 
$P({\bm{k}^N})=\sum_{j=1}^{N}k_j$ and $E({\bm{k}}^N)=\sum_{j=1}^{N}k_j^2$, respectively. 
If we specify a set of the Bethe quantum numbers 
$I_1<\cdots<I_N$, the BAEs \eqref{BAE} have a unique real solution $k_1 < \cdots < k_N$.

We consider a quantum quench, which mimics a coherent splitting of the 1D Bose gas \cite{Gring2012}.
The splitting process is illustrated in Fig.~\ref{setupLL}, where the Lieb-Liniger model is split into two parts (left and right) that do not interact with each other.
This process places each particle in a superposition of the left and right states.
Let us introduce the field operators of ``symmetric bosons'' $\hat{\psi}_s(x)=(\hat{\psi}_{\rm L}(x)+\hat{\psi}_{\rm R}(x))/{\sqrt{2}}$ and ``anti-symmetric bosons'' $\hat{\psi}_a(x)=(\hat{\psi}_{\rm L}(x)-\hat{\psi}_{\rm R}(x))/{\sqrt{2}}$, where $\hat{\psi}_{\rm L}(x)$ and $\hat{\psi}_{\rm R}(x)$ are operators of different Bose fields.
Before the split, symmetric bosons correspond to the ground state in the $y$-axis (the direction perpendicular to the 1D Bose gas), while anti-symmetric bosons are responsible for the excitation in the  $y$-axis.
If the angular frequency of the radial trapping potential is denoted by $\omega_{\perp}$, which is assumed to be sufficiently large compared with any other relevant energy scales, this excitation energy is given by $\omega_{\perp}$.
Thus the initial Hamiltonian before the split, $\hat{H_{0}}$, is given by
\begin{align}
\hat{H}_0=&\int_{0}^{L}dx\left({\partial_x}\hat{\psi}_s^{\dagger}(x){\partial_x}\hat\psi_s(x)
\right.\nonumber \\
&+c \hat\psi_s^{\dagger}(x)\hat\psi_s^{\dagger}(x)\hat\psi_s(x)\hat\psi_s(x)
\left. +\omega_{\perp}\hat\psi_a^{\dagger}(x)\hat\psi_a(x)\right).
\end{align}
After the split, the left gas does not interact with the right one, so
the Hamiltonian after the quench $H_{1}$ is given by $\hat{H}_1=\hat{H}_{\rm L}+\hat{H}_{\rm R}$, where
\begin{equation}
\begin{split}
\hat{H}_{\rm L/R}&={\int}_{0}^{L}dx(\partial_x\hat{\psi}_{\rm L/R}^{\dagger}(x)\partial_x\hat{\psi}_{\rm L/R}(x)\\
&+c'\hat{\psi}_{\rm L/R}^{\dagger}(x)\hat{\psi}_{\rm L/R}^{\dagger}(x)\hat{\psi}_{\rm L/R}(x)\hat{\psi}_{\rm L/R}(x)).
\end{split}
\end{equation}
The coupling constant after the quench $c'$ is chosen to be $c'=c/2$.
We note that the number of particles on the left or on the right obeys a binomial distribution since each particle is superposed between the left and right states with equal weights.
We denote the numbers of particles of the left and right by $M$ and $N-M$, respectively.

Our initial state $|\Psi(0)\rangle$ is chosen as the ground state of $H_0$, and the state at time $t$ is given by $|\Psi(t)\rangle=e^{-i{\hat{H}}_1t}|\Psi(0)\rangle$.
By expanding $|\Psi(0)\rangle$ in the basis of eigenstates of $\hat{H}_1$, we have 
\begin{equation}
\begin{split}
|\Psi(t)\rangle=\sum_{M=0}^{N}\sum_{\bm{k}_{\rm L}^M,\bm{k}_{\rm R}^{N-M}}^{\infty}&C(\bm{k}_{\rm L}^M,\bm{k}_{\rm R}^{N-M})e^{-iE (\bm{k}_{\rm L}^M,\bm{k}_{\rm R}^{N-M})t}\\
&\times |\bm{k}_{\rm L}^M\rangle|\bm{k}_{\rm R}^{N-M}\rangle,
\end{split}
\label{eq:state}
\end{equation}
where
\begin{equation}
\hat{H}_1|{\bm{k}_{\rm L}}^M\rangle|{\bm{k}}_{\rm R}^{N-M}\rangle=E ({\bm{k}}_{\rm L}^M,{\bm{k}}_{\rm R}^{N-M})|{\bm{k}_{\rm L}}^M\rangle|{\bm{k}}_{\rm R}^{N-M}\rangle
\end{equation}
with $E ({\bm{k}}_{\rm L}^M,{\bm{k}}_{\rm R}^{N-M}) \equiv E ({\bm{k}}_{\rm L}^M)+E ({\bm{k}}_{\rm R}^{N-M})$.
The time evolution is easily calculated once we determine the expansion coefficients $\{C(\bm{k}_{\rm L}^M,\bm{k}_{\rm R}^{N-M})\}$ since $E (\bm{k}_{\rm L}^M, \bm{k}_{\rm R}^{N-M})$ can be calculated exactly by the Bethe ansatz method.

To examine whether our system displays prethermalization for several numbers of particles, we calculate two-point correlation functions in Fig.~\ref{correlation}.
In Fig.~\ref{correlation}(a), the red curve shows the result of $\overline{C_{\rm L}}(x)$ which we calculate by using the Bethe ansatz method.
The blue curve shows the thermal equilibrium average $C_{\rm L}^{\rm eq}(x)={\rm{Tr}}\rho_{\rm{{can}}}\hat{\psi}_{\rm L}^{\dagger}(x)\hat{\psi}_{\rm L}(0)$.
The autocorrelation function $\overline{C_{\rm L}}(x)$ is well described by the equilibrium distribution $C_{\rm L}^{\rm eq}(x)$ if $\beta$ is appropriately determined, e.g. $\beta=0.185$ for $N=4$.

In contrast, in Fig.~\ref{correlation}(b),
the cross-correlation function $\overline{C_{\rm LR}}(x)$ depicted by the red curve is compared with the equilibrium correlation function $C_{\rm LR}^{\rm eq}(x)$ depicted by the blue curve, where 
$C_{\rm LR}^{\rm eq}(x) = {\rm{Tr}}\rho_{\rm can}\hat{\psi}^{\dagger}_{\rm L}(x)\hat{\psi}_{\rm L}(0)\hat{\psi}^{\dagger}_{\rm R}(0)\hat{\psi}_{\rm R}(x)$.
At short distances, they agree with each other at a temperature determined by the auto-correlation described above.
However, they disagree at long distances.
In other words, the left-right interference modes with long wavelength have temperatures different from the other modes.
In a true equilibrium state, all the modes should have the same temperature, and hence this result shows that the system is thermalized only partially, which is a characteristic of prethermalization.

Now, let us explain the physical mechanism of the prethermalization.
Let  $\bm{k}_{\rm L}^M$ and $\bm{k}_{\rm R}^{N-M}$ be the Bethe momenta of the left and the right, respectively.
It follows from the parity symmetry that $E ({\bm{k}}_{\rm L}^M,{\bm{k}}_{\rm R}^{N-M})=E (-{\bm{k}}_{\rm L}^M,-{\bm{k}}_{\rm R}^{N-M})$ and $C ({\bm{k}}_{\rm L}^M,{\bm{k}}_{\rm R}^{N-M})=C (-{\bm{k}}_{\rm L}^M,-{\bm{k}}_{\rm R}^{N-M})$.
By neglecting the other type of degeneracies, the infinite-time average of the density matrix  $\overline{\rho}=\overline{|\Psi(t)\rangle\langle\Psi(t)|}$ is given by
\begin{equation}
\overline{\rho} \simeq \sum_{{\bm{k}}_{\rm L}^M,{\bm{k}}_{\rm R}^{N-M}}|C({\bm{k}}_{\rm L}^M,{\bm{k}}_{\rm R}^{N-M})|^2| \Phi_{{\bm{k}}^N}\rangle\langle \Phi_{{\bm{k}}^N}|,
\end{equation}
where 
\begin{equation}
|\Phi_{\bm{k}^N}\rangle=\frac{1}{\sqrt{2}}(|{\bm{k}}_{\rm L}^M,{\bm{k}}_{\rm R}^{N-M}\rangle + |-{\bm{k}}_{\rm L}^M,-{\bm{k}}_{\rm R}^{N-M}\rangle)
\end{equation}
is a fully entangled state consisting of the two degenerate states.
The diagonal components in the basis $\{|{\bm{k}}_{\rm L}^M, {\bm{k}}_{\rm R}^{N-M}\rangle \}$ are given by 
\begin{equation}
\begin{split}
\rho_{\rm d}& \simeq \sum_{{\bm{k}}_{\rm L}^M,{\bm{k}}_{\rm R}^{N-M}}|C({\bm{k}}_{\rm R}^M,{\bm{k}}_{\rm L}^{N-M})|^2
\left( |{\bm{k}}_{\rm L}^M,{\bm{k}}_{\rm R}^{N-M}\rangle\langle{\bm{k}}_{\rm L}^M,{\bm{k}}_{\rm R}^{N-M}| \right. \\
&\left. +|-{\bm{k}}_{\rm L}^M,-{\bm{k}}_{\rm R}^{N-M}\rangle\langle-{\bm{k}}_{\rm L}^M,-{\bm{k}}_{\rm R}^{N-M}|\right),
\end{split}
\end{equation}
and the off-diagonal components are given by
\begin{equation}
\begin{split}
\rho_{\rm {off-d}}& \simeq \sum_{{\bm{k}}_{\rm L}^M,{\bm{k}}_{\rm R}^{N-M}}|C({\bm{k}}_{\rm R}^M,{\bm{k}}_{\rm L}^{N-M})|^2\\
&\times \left(|{\bm{k}}_{\rm L}^M,{\bm{k}}_{\rm R}^{N-M}\rangle\langle{-\bm{k}}_{\rm L}^M,-{\bm{k}}_{\rm R}^{N-M}| \right.\\
&\left. +|-{\bm{k}}_{\rm L}^M,-{\bm{k}}_{\rm R}^{N-M}\rangle\langle{\bm{k}}_{\rm L}^M,{\bm{k}}_{\rm R}^{N-M}| \right),
\end{split}
\end{equation}
which arise from the initial entanglement between the left and right.
Because of the entanglement of ${|\Phi_{\bm{k}^N}\rangle}$, the off-diagonal elements $\rho_{\rm {off-d}}$ do not contribute to the expectation value of any observable involving the left or right alone such as ${\rm Tr}\overline{\rho} \hat{O}_{\rm L}={\rm Tr}{\rho_d} \hat{O}_{\rm L}$ for any observable $\hat{O}_{\rm L}$ consisting of $\hat{\psi}_{\rm L}(x)$ and $\hat{\psi}_{\rm L}^{\dagger}(x)$.
In contrast, when we consider the interference between  the left and the right, off-diagonal components have a nonvanishing contribution and prevent thermalization.

The crucial point is that the quantum coherence between the degenerate states, $|\bm{k}_{\rm L}^M,\bm{k}_{\rm R}^{N-M}\rangle$ and $|-\bm{k}_{\rm L}^M,-\bm{k}_{\rm R}^{N-M}\rangle$, do not vanish even after the infinite-time average, and hence the initial entanglement between the left and the right can survive, albeit partially.
This is a general result in quantum mechanics and not restricted to the Lieb-Liniger model.
We can indeed show that if $H_{\rm L}$ and $H_{\rm R}$ are degenerate, the initial entanglement between the left and the right can significantly affect the steady state.
An eigenstate of $H_{\rm L}+H_{\rm R}$ is denoted by $|n,i;m,j\rangle$, where $H_{\rm L}|n,i;m,j\rangle=E_n|n,i;m,j\rangle$ and $H_{\rm R}|n,i;m,j\rangle=E_m|n,i;m,j\rangle$.
The indices $i=1,2,\dots,d_n$ and $j=1,2,\dots,d_m$ distinguish the degenerate states.
Then the infinite-time average of the density matrix reads
\begin{equation}
\bar{\rho}=\sum_{n\leq m}p_{nm}|\Phi_{nm}\rangle\langle\Phi_{nm}|,
\end{equation}
where $p_{nm}$ is the probability that the energy of the total system is given by $E_n+E_m$ and $|\Phi_{nm}\rangle$ is the projection of the initial state onto the subspace spanned by the energy eigenstates with the energy $E_n+E_m$.
If $|\Phi_{nm}\rangle$ has entanglement, $\bar{\rho}$ also has entanglement in general.
Whether or not $|\Phi_{nm}\rangle$ has entanglement depends, of course, on the specific problem, but the above argument shows that the initial entanglement can affect the steady state if there are energy degeneracies.
See Supplementary Material for more detail.
%
\begin{figure}[t]
\includegraphics[width=0.9\columnwidth]{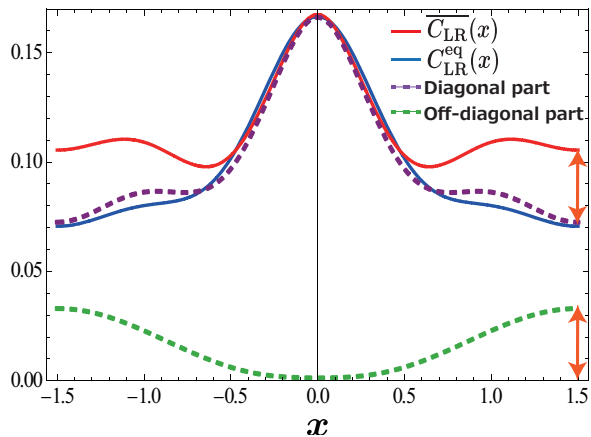}
\caption{{\bf{Diagonal and off-diagonal components in the infinite-time average of the density matrix $\overline\rho(t)$.}}
The diagonal and off-diagonal components are added to Fig.~\ref{correlation} for the case of $N=3$.
The purple dashed curve and green dashed curve show the diagonal and off-diagonal contributions, respectively.
The lengths of the orange arrows are equal.}
\label{diagnal}
\end{figure}

We numerically calculate the diagonal and off-diagonal components as shown in Fig.~\ref{diagnal}.
The off-diagonal contribution emerges at long distances, which gives a major contribution to the deviation of $\overline{C_{\rm LR}}(x)$ from its equilibrium counterpart as shown in Fig.~\ref{correlation}(b).
On the other hand, the contribution from the diagonal elements agrees with the equilibrium curve.
Thus, the nonlocal correlation between the left and the right plays a decisive role for the entanglement prethermalization.

Finally, we propose an experimental setup to observe entanglement prethermalization,
that is, a
coherent splitting of an ultracold interacting Bose/Fermi gas
in a 3D spherical harmonic trap into halves.
As discussed above,
entanglement prethermalization occurs
if $H_{\rm L}$ and $H_{\rm R}$ are degenerate,
and this condition holds in this setup due to the 3D-rotational invariance
if the interactions between particles are invariant under 3D rotations.
Now that a cigar-shaped-trap version of the setup has been realized
with small numbers of fermions~\cite{Serwane2011},
the setup can be realized by using equal trapping frequencies along all the three directions.
Since the degeneracy is lifted in the cigar-shaped trap
and entanglement prethermalization is not present in there,
it is worthwhile investigating how entanglement prethermalization
is affected under the continuous deformation of the trap
from the spherical one to the cigar-shaped one.

In conclusion, we have discovered a new mechanism of prethermalization by calculating the infinite-time average of two-point correlation functions in a few-body system; the initial entanglement between the two spatially separated subsystems plays a crucial role in the prethermalization observed in a coherently split 1D Bose gas.
The diagonal elements of the time-averaged density matrix stand for coherence within one subsystem which explains the equilibrium correlation functions at an effective temperature.
On the other hand, the effect of entanglement manifests itself in the off-diagonal elements, 
and it cannot be described by the Gibbs ensemble at any effective temperature.
Such an entanglement prethermalization is found in a few-body system and should be observed in experiments.
Our study reveals that prethermalization found in coherently split systems belongs to a new class of prethermalization, in which the initial entanglement plays a crucial role.
The existence of energy degeneracies in each of the left and the right subsystems is crucial and the mechanism of the entanglement prethermalization does not depend on the dimensionality, the quantum statistics, or the integrability.
Therefore, the entanglement prethermalization should occur not only in the Lieb-Liniger model but also in more general systems such as a system composed of three dimensional harmonic oscillators interacting with each other.

\bigskip
\noindent
{\bf Methods}

{\small
\noindent
{\bf How to obtain expansion coefficients.}
The expansion coefficients $\{ C(\bm{k}_{\rm L}^M,\bm{k}_{\rm R}^{N-M})\}$ in equation ~(\ref{eq:state}) are defined as $C(\bm{k}_{\rm L}^M,\bm{k}_{\rm R}^{N-M})=\langle\bm{k}_{\rm L}^M,\bm{k}_{\rm R}^{N-M}|\Psi(0)\rangle$, where $|\Psi(0)\rangle$ is the ground state of $\hat{H}_0$.
To obtain them, we calculate the matrix elements of $\hat{H}_0$ in the basis $\{|\bm{k}_{\rm L}^M,\bm{k}_{\rm R}^{N-M}\rangle\}$ and diagonalize this matrix.
Each matrix element $\langle{\bm{k}'}_{\rm L}^{M'},{\bm{k}'}_{\rm R}^{N-M'}|\hat{H}_0|\bm{k}_{\rm L}^M,\bm{k}_{\rm R}^{N-M}\rangle$ contains matrix elements of the field operators such as $\langle{\bm{k}'}_{\rm L}^{M-1}|\hat{\psi}_{\rm L}(0)|\bm{k}_{\rm L}^M\rangle$ and $\langle{\bm{k}'}_{\rm L}^M|\hat{\psi}^{\dagger}_{\rm L}(0)\hat{\psi}_{\rm L}(0)|\bm{k}_{\rm L}^M\rangle$.
These matrix elements can be evaluated by using the Gaudin-Korepin \cite{Gaudin1983, Korepin86} and Slavnov formulae \cite{Slavnov1989, Slavnov1990, Caux2007}.
A difficulty here is that the dimension of the matrix $\hat{H}_1$ is infinite because the Bethe quantum numbers $\{I_j\}$ run over all the integers or all the half-integers depending on whether the number of bosons is odd or even, respectively.
We therefore truncate the range of Bethe quantum numbers as $-I_{\rm max}\leq I_j\leq I_{\rm max}$ for all $j$, where the value of $I_{\rm max}$ is chosen sufficiently large.
We emphasize that this is just an approximation for the initial state and not for the dynamics.
Because eigenenergies for the Bethe eigenstates can be evaluated in a numerically exact manner by solving the Bethe ansatz equation, the dynamics is still exact.
We can then diagonalize the truncated matrix $\hat{H}_1$, and the eigenvector associated with the minimum eigenvalue gives us the approximate expansion coefficients $\{C(\bm{k}_{\rm L}^M,\bm{k}_{\rm R}^{N-M})\}$.

\noindent
{\bf Equilibrium correlation functions.}
One might think that the equilibrium density matrix associated with $\hat{H}_1=\hat{H}_{\rm L}+\hat{H}_{\rm R}$ is given by $\exp[-\beta(\hat{H}_{\rm L}+\hat{H}_{\rm R})]/{\rm Tr}\exp[-\beta(\hat{H}_{\rm L}+\hat{H}_{\rm R})]$.
In our setup, however, there are some conserved quantities which we must take into account.
Firstly, the total momentum is conserved.
Since the initial state is the ground state of $\hat{H}_0$, the total momentum is zero:
$P(\bm{k}_{\rm L}^M)+P(\bm{k}_{\rm R}^{N-M})=0$.
Secondly, as each boson is distributed to the left or right randomly and the redistribution of particles do not take place after the quench, the number of particles $M$ in the left Bose gas obeys the binomial distribution, $2^{-N}N!/M!(N-M)!$.
The principle of maximum entropy under the above constraints lead us to the equilibrium density matrix of the form,
\begin{align}
\rho_{\rm can}=\sum_{M=0}^N\frac{1}{2^N}\frac{N!}{M!(N-M)!}\sum_{\bm{k}_{\rm L}^M,\bm{k}_{\rm R}^{N-M}} \delta_{P(\bm{k}_{\rm L}^M)+P(\bm{k}_{\rm R}^{N-M}),0}
\nonumber \\
\times e^{-\beta(E(\bm{k}_{\rm L}^M)+E(\bm{k}_{\rm R}^{N-M}))} |\bm{k}_{\rm L}^M,\bm{k}_{\rm R}^{N-M}\rangle \langle\bm{k}_{\rm L}^M,\bm{k}_{\rm R}^{N-M}|.
\label{eq:rho_eq}
\end{align}
In the calculation of equilibrium correlation functions given in this work, equation ~(\ref{eq:rho_eq}) is used.
}

\noindent
{\bf{Acknowledgements}}

{\small
\noindent
We would like to thank T. Deguchi and N. Sakumichi for discussions.
This work was supported by
KAKENHI Grant No. 26287088 from the Japan Society for the Promotion of Science, 
a Grant-in-Aid for Scientific Research on Innovation Areas ``Topological Quantum Phenomena'' (KAKENHI Grant No. 22103005),
the Photon Frontier Network Program from MEXT of Japan,
and the Mitsubishi Foundation.
E.K. acknowledges support from Institute for Photon Science and Technology.
T.N.I. acknowledges the JSPS for financial support (Grant No. 248408).
}   

\noindent
{\bf{Author contributions}}

{\small
\noindent
All authors contributed to the work of this manuscript.
}

\noindent
{\bf{Competing financial interests}}

{\small
\noindent
The authors declare no competing financial interests.
}

\newpage
\onecolumngrid
\section*{Supplemental Material: General argument on the mechanism of entanglement prethermalization}

Here we discuss the generality of the mechanism of entanglement prethermalization.
As discussed in the main text, the entanglement prethermalization results from the long-lasting influence of the initial entanglement between the ``left'' and ``right'' subsystems.
Usually, the effect of the initial entanglement fades out during the time evolution and is not relevant for long-time behavior of physical quantities.
If there are some energy degeneracies, however, the initial entanglement remains and can give large contribution to the measured values of physical quantities.

We consider the Hamiltonian $H=H_{\rm L}+H_{\rm R}$, where $H_{\rm L}$ and $H_{\rm R}$ are the Hamiltonians of the left and right subsystems, respectively, and commute with each other.
We assume that $H_{\rm L}$ and $H_{\rm R}$ have the same energy spectrum, i.e. the two subsystems are identical.
The energy eigenstate of $H$ is denoted by $|n,i;m,j\rangle$, where $H_{\rm L}|n,i;m,j\rangle=E_n|n,i;m,j\rangle$ and $H_{\rm R}|n,i;m,j\rangle=E_m|n,i;m,j\rangle$.
The degree of degeneracies of the eigenstates of $H_{\rm L/R}$ with energy $E_n$  is denoted by $d_n$, and thus $i=1,2,\dots, d_n$ and $j=1,2,\dots, d_m$.
We assume the non-resonant condition; if $E_n-E_m=E_k-E_l\neq 0$, $n=k$ and $m=l$, or equivalently, if $E_n+E_m=E_k+E_l$, $n=k$ and $m=l$ or $n=l$ and $m=k$.

From this assumption, the infinite-time average of the density matrix is given by
\begin{equation}
\bar{\rho}\equiv \lim_{T\rightarrow\infty}\frac{1}{T}\int_0^T|\Psi(t)\rangle\langle\Psi(t)|dt
=\sum_{n\leq m}\mathcal{P}_{nm}|\Psi(0)\rangle\langle\Psi(0)|\mathcal{P}_{nm},
\end{equation}
where $\mathcal{P}_{nm}$ is the projection onto the Hilbert subspace with the energy $E_n+E_m$, that is,
\begin{equation}
\mathcal{P}_{nm}\equiv\left\{
\begin{aligned}
&\sum_{i=1}^{d_n}\sum_{j=1}^{d_m}\left(|n,i;m,j\rangle\langle n,i;m,j|+|m,j;n,i\rangle\langle m,j;n,i|\right) \qquad (n\neq m), \\
&\sum_{i,j=1}^{d_n}|n,i;n,j\rangle\langle n,i;n,j| \qquad (n=m).
\end{aligned}
\right.
\end{equation}
By defining $|\Phi_{nm}\rangle\equiv\mathcal{P}_{nm}|\Psi(0)\rangle/\sqrt{\langle\Psi(0)|\mathcal{P}_{nm}|\Psi(0)\rangle}$, $\bar{\rho}$ is expressed as
\begin{equation}
\bar{\rho}=\sum_{n\leq m}p_{nm}|\Phi_{nm}\rangle\langle\Phi_{nm}|
\end{equation}
with $p_{nm}=\langle\Psi(0)|\mathcal{P}_{nm}|\Psi(0)\rangle$.
If we define $C_{(n,i),(m,j)}=\langle n,i;m,j|\Psi(0)\rangle$, the explicit form of $|\Phi_{nm}\rangle$ is given by
\begin{equation}
|\Phi_{nm}\rangle\propto\mathcal{P}_{nm}|\Psi(0)\rangle
=\sum_{i=1}^{d_n}\sum_{j=1}^{d_m}\left(C_{(n,i),(m,j)}|n,i;m,j\rangle+C_{(m,j),(n,i)}|m,j;n,i\rangle\right)
\label{eq:Phi}
\end{equation}
for $n\neq m$, and 
\begin{equation}
|\Phi_{nn}\rangle\propto\mathcal{P}_{nn}|\Psi(0)\rangle
=\sum_{i,j=1}^{d_n}C_{(n,i),(n,j)}|n,i;n,j\rangle.
\end{equation}
Although the coherence between the energy eigenstates with different energies is lost by the time averaging, the coherence among the degenerate states remains and the left subsystem is still entangled with the right one in general.
This entanglement contributes to a significant deviation from the thermal equilibrium.

When the initial state has the parity symmetry, $C_{(n,i),(m,j)}=C_{(m,j),(n,i)}$,
\begin{equation}
|\Phi_{nm}\rangle\propto\sum_{i=1}^{d_n}\sum_{j=1}^{d_m}C_{(n,i),(m,j)}\frac{|n,i;m,j\rangle+|m,j;n,i\rangle}{\sqrt{2}}.
\label{eq:Phi_parity}
\end{equation}
If a coefficient $C_{(n,i),(m,J)}$ is not decoupled like $C_{(n,i),(m,j)}\neq C_{n,i}C_{m,j}$, the left subsystem and the right subsystem are entangled.
The initial entanglement remains even after the time average when there are degeneracies.

We note that even if there is no initial entanglement between the left and the right and the time evolutions of the two subsystems are completely independent, some quantum correlations may appear in the time-averaged density matrix, as is obvious from Eq.~(\ref{eq:Phi_parity}); even if there is no initial entanglement and $C_{(n,i),(m,j)}=C_{(n,i)}C_{(m,j)}$, $|\Phi_{nm}\rangle$ is an entangled state.
In order to understand this aspect, we consider a two-level system, $n,m=1,2$ and $d_1=d_2=1$, and simply write $|n,1;m,1\rangle =|n,m\rangle=|n\rangle_{\rm L}\otimes|m\rangle_{\rm R}$.
Here $H_{\rm L/R}|0\rangle_{\rm L/R}=0$ and $H_{\rm L/R}|1\rangle_{\rm L/R}=\varepsilon|1\rangle_{\rm L/R}$.
The initial state is assumed to be
$$
|\Psi(0)\rangle=\frac{|0,0\rangle+|0,1\rangle+|1,0\rangle+|1,1\rangle}{2}=\frac{|0\rangle_{\rm L}+|1\rangle_{\rm L}}{\sqrt{2}}\otimes\frac{|0\rangle_{\rm R}+|1\rangle_{\rm R}}{\sqrt{2}}.
$$
The infinite-time average of the density matrix is then given by
$$
\bar{\rho}=\frac{1}{4}(|0,0\rangle+|1,1\rangle)+\frac{1}{2}\frac{|0,1\rangle+|1,0\rangle}{\sqrt{2}}\frac{\langle 0,1|+\langle 1,0|}{\sqrt{2}}.
$$
Actually, this state does not have entanglement because $\bar{\rho}$ is separable by definition:
$$
\bar{\rho}=\frac{\varepsilon}{2\pi}\int_0^{2\pi/\varepsilon}\left(\frac{|0\rangle_{\rm L}+e^{-i\varepsilon t}|1\rangle_{\rm L}}{\sqrt{2}}\frac{\langle0|_{\rm L}+e^{i\varepsilon t}\langle 1|_{\rm L}}{\sqrt{2}}\right)\otimes\left(\frac{|0\rangle_{\rm R}+e^{-i\varepsilon t}|1\rangle_{\rm R}}{\sqrt{2}}\frac{\langle0|_{\rm R}+e^{i\varepsilon t}\langle 1|_{\rm R}}{\sqrt{2}}\right)dt.
$$
However, it has nonzero quantum discord \cite{Ollivier2001}, and hence the correlation between the left and the right in $\bar{\rho}$ is not fully classical even in this case.

When a physical quantity of interest $O$ has no large matrix element of the form $\langle n,i;m,j|O|m,j';n,i'\rangle$, we can neglect the quantum coherence between $|n,i;m,j\rangle$ and $|m,j';n,i'\rangle$ in $\bar{\rho}$.
In that case, we can use the projection operator $\mathcal{P}'_{nm}$ instead of $\mathcal{P}_{nm}$ defined by
\begin{equation}
\mathcal{P}'_{nm}=\sum_{i=1}^{d_n}\sum_{j=1}^{d_m}|n,i;m,j\rangle\langle n,i;m,j|
\end{equation}
for $n\neq m$ and $\mathcal{P}'_{nn}=\mathcal{P}_{nn}$.
The infinite-time average of the density matrix is then approximated by 
\begin{equation}
\bar{\rho}\approx\sum_{n,m}\mathcal{P}'_{nm}|\Psi(0)\rangle\langle\Psi(0)|\mathcal{P}'_{nm}
=\sum_{n,m}p'_{nm}|\Phi'_{nm}\rangle\langle\Phi'_{nm}|
\end{equation}
in the sense of ${\rm Tr}\bar{\rho}O\approx{\rm Tr}\bar{\rho}'O$.
Here $p'_{nm}=\langle\Psi(0)|\mathcal{P}'_{nm}|\Psi(0)\rangle$ and
\begin{equation}
|\Phi'_{nm}\rangle\propto\mathcal{P}'_{nm}|\Psi(0)\rangle
=\sum_{i=1}^{d_n}\sum_{j=1}^{d_m}C_{(n,i;m,j)}|n,i;m,j\rangle.
\end{equation}

In the case of the Lieb-Liniger model discussed in the main text, $n$ and $m$ correspond to $\bm{k}_{\rm L}^M$ and $\bm{k}_R^{N-M}$, respectively.
The degeneracy is due to the reflection symmetry, $\bm{k}_L^M\rightarrow -\bm{k}_L^M$, so $d_n=2$ for every $n$.
Therefore, 
\begin{equation}
\begin{aligned}
|n,1;m,1\rangle&\rightarrow |\bm{k}_{\rm L}^M,\bm{k}_{\rm R}^{N-M}\rangle, \\
|n,1;m,2\rangle&\rightarrow |\bm{k}_{\rm L}^M,-\bm{k}_{\rm R}^{N-M}\rangle, \\
|n,2;m,1\rangle&\rightarrow |-\bm{k}_{\rm L}^M,\bm{k}_{\rm R}^{N-M}\rangle, \\
|n,2;m,2\rangle&\rightarrow |-\bm{k}_{\rm L}^M,-\bm{k}_{\rm R}^{N-M}\rangle. \\
\end{aligned}
\end{equation}
In the initial state discussed in the main text, the total momentum is zero, $P(\bm{k}_{\rm L}^M)+P(\bm{k}_{\rm R}^{N-M})=0$.
Therefore, $C_{(n,1),(m,2)}=C_{(n,2),(m,1)}=0$, and $C_{(n,1),(m,1)}=C_{(n,2),(m,2)}$ because of the reflection symmetry. 
Thus 
\begin{equation}
|\Phi'_{nm}\rangle=\frac{|\bm{k}_{\rm L}^M,\bm{k}_{\rm R}^{N-M}\rangle+|-\bm{k}_{\rm L}^M,-\bm{k}_{\rm R}^{N-M}\rangle}{\sqrt{2}},
\end{equation}
and clearly the left and the right subsystems are entangled.

In conclusion, in the general setup, we have shown that the infinite-time average of the density matrix can have some amount of entanglement between the left subsystem and the right one. 
The only requirement for this is that there are degeneracies in $H_{\rm L/R}$; we have not assumed the integrability of the system, the Bose statistics, the dimensionality, and so on.
The entanglement prethermalization is therefore considered to be a generic phenomenon that can be observed in split subsystems and is not restricted to the one-dimensional integrable Bose gas.

\end{document}